\documentclass[showpacs,twocolumn,prl,floatfix,aps]{revtex4}
\usepackage{graphicx}

\begin{document}

\hsize\textwidth\columnwidth\hsize\csname@twocolumnfalse\endcsname

\title{Effects of the Dzyaloshinskii-Moriya interaction on low
energy magnetic excitations in copper benzoate}

\author{J. Z. Zhao$^1$, X. Q. Wang$^{1,2,3}$, T. Xiang$^{1,2}$,
Z. B. Su$^{1,2}$, and L. Yu$^{1,2}$}

\address{$^1$Institute of Theoretical Physics, CAS, Beijing 100080
China}

\address{$^2$The Interdisciplinary Center of Theoretical Studies, CAS,
Beijing 100080, China}

\address{$^3$Institute for Solid State Physics, University of Tokyo,
Kashiwa, Chiba 277-8581, Japan}

\begin{abstract}
We have investigated the physical effects of the
Dzyaloshinskii-Moriya (DM) interaction in copper benzoate.
In the low field limit,
the spin gap is found to vary as $H^{2/3}\ln ^{1/6}(J/\mu_BH_s)$
($H_s$: an effective staggered field induced by the external field $H$)
in agreement with the prediction of conformal field theory, while the staggered
magnetization varies as $H^{1/3}$ and the $\ln^{1/3}(J/\mu_BH_s)$
correction predicted by conformal field theory is not confirmed.
The linear scaling behavior between the momentum shift 
and the magnetization is broken. We have determined the coupling
constant of the DM interaction and have given a
complete quantitative account for the field dependence of the spin
gaps along all three principal axes, without resorting to
additional interactions like interchain coupling. A crossover to
strong applied field behavior is predicted for further experimental
verification.
\end{abstract}


\maketitle

Recently some novel magnetic properties were discovered in a
variety of quasi-one dimensional materials, such as copper
benzoate Cu(C$_6$D$_5$COO)$_2$3D$_2$O\cite{Den1,Den2},
Yb$_4$As$_3$ \cite{Kohgi,Fulde,Osh2} and
BaCu$_2$Si$_2$O$_7$\cite{Tsukada}. In these materials, the
Dzyaloshinskii-Moriya (DM) interaction\cite{Dyz,Mor,Osh1} plays an
important role, especially in an applied magnetic field. This has
stimulated extensive investigation on the physical properties of
the DM interaction. However, this interaction is rather difficult
to handle analytically, which has brought much uncertainty in the
interpretation of experimental data and has limited our
understanding of many interesting quantum phenomena of
low-dimensional magnetic materials.

For copper benzoate, Dender {\it et al}\cite{Den1,Den2}  found
that the spin excitation gap shows a peculiar field dependence,
$\Delta \sim H^{0.65}$, in low fields. On the contrary,
excitations remain gapless in the S=1/2 Heisenberg model below a
critical field. Oshikawa and Affleck (OA) suggested that this
field dependence of the gap is due to a staggered magnetic field
induced by the DM interaction in addition to the staggered
g-factor in a uniform field\cite{mutter,Osh1}. However, a
satisfactory explanation for the field-dependence of the energy
gaps in all three directions is still lacking\cite{Ess,lou}. It
was argued that the inconsistency between the experimental data
and theoretical results might be due to the neglect of the
interchain coupling and/or anisotropic interaction terms in the
low-field effective model used by Oshikawa and Affleck
\cite{Osh1,Ess}. We believe this issue can be clarified by a
thorough study of the DM interaction and a direct comparison with
experiments.

Copper benzoate is a quasi-1D spin-1/2 antiferromagnetic
Heisenberg system. The chain direction is the c-axis. It contains
two types of alternating and slightly tilted CuO$_8$ octahedra.
This leads to two inequivalent Cu$^{++}$ ions and an alternating
DM coupling \cite{Date}. In an applied field, copper
benzoate can be modeled by the following Hamiltonian,
\begin{eqnarray}
\hat{H} = \sum_i&& \left( J\hat{\bf S}_i\cdot \hat{{\bf
S}}_{i+1}-(-)^i{\bf D} \cdot \hat{\bf S}_i\times \hat{\bf
S}_{i+1}\right. \nonumber \\
&& \left. -\mu_B{\bf H}\cdot \left[{\bf g}^u+(-)^i{\bf
g}^s\right]\cdot \hat{\bf S}_i \right)
 \label{HSDMZ}
\end{eqnarray}
where the three terms in the summation are the antiferromagnetic
Heisenberg, DM and Zeeman splitting interactions, respectively.
The exchange coupling constant $J$, determined from the neutron
scattering measurements, is about $1.57{\rm meV}$. The DM
interaction is much weaker than the Heisenberg term. The ${\bf
D}$-vector, primarily aligned along the $a^{\prime \prime }$ axis,
will be determined numerically. ${\bf g}^u$ and ${\bf g}^s$ are
the uniform and staggered components of the alternating g-tensor
as given in Ref. \onlinecite{Date} for copper benzoate.

The DM term can in principle be eliminated by performing a spin
rotation about the ${\bf D}$ vector by an angle $\alpha=\pm \frac
1 2 \tan^{-1} (D/J)$ on the alternating sites. This results in a
small exchange anisotropy and an effective staggered magnetic
field in addition to the ${\bf g}^s$-term in (\ref{HSDMZ}). The
total effective staggered field is approximately given by
\begin{equation}
{\bf H}_s \approx \left( {\bf g}^s - {1\over 2J} {\bf D}\times
{\bf g}^u \right) \cdot {\bf H} \label{stagger}
\end{equation}
up to the leading order in $D/J$. In Ref. \onlinecite{Osh1}, OA
studied the isotropic Heisenberg model with a staggered field
after eliminating the DM term and neglecting all anisotropic
terms. They estimated that the DM coupling constant is about ${\bf
D} =(0.13,0.0,0.02)J$ from the specific heat, neutron scattering
and ESR measurement data\cite{Osh1,Den1}.

In order to explore the low-energy properties of copper benzoate,
we use density matrix renormalization group (DMRG)
\cite{white,peschel} to study directly the Hamiltonian defined by
Eq. (1). Since neither the total spin nor its z-component is a
good quantum number, the calculation is computer time consuming.
Open boundary conditions are used and up to 400 states are
retained in our calculation. The convergence of the results is
systematically examined and the truncation error is less than
$10^{-7}$. To obtain the values of gaps and magnetizations in the
thermodynamic limit, an extrapolation from the DMRG calculations
up to 1000 sites is done.

\begin{figure}[h]
\hspace{-0.2cm}{\includegraphics[width=0.39\textwidth]{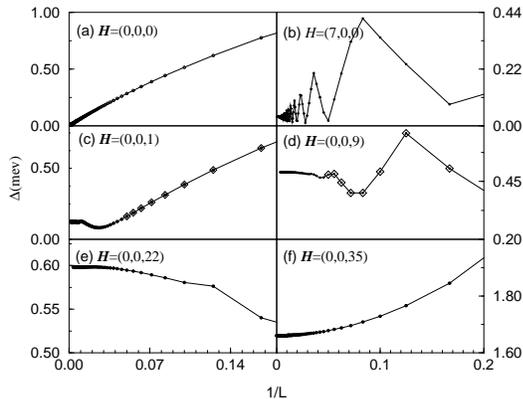}}
\vspace{-0.1cm}
\caption{Spin gaps versus 1/L in different applied
fields for ${\bf D}=(0.13,0.0,0.02)J$. The exact diagonalization results
($\diamond$) for small lattices are also shown for comparison.}
\label{fig1}
\end{figure}

Figure \ref{fig1} shows the spin excitation gap as a function of
inverse lattice size $L$ in different applied fields $H$. In zero
field, there is no gap in the excitation spectrum in the
thermodynamic limit and the gap decreases monotonically with $L$
increasing. However, in finite fields, the size dependence of the
gap becomes more complicated and changes dramatically with $H$
increasing. The oscillations of the gaps are due to the competition
between the uniform and staggered magnetic fields. The most
pronounced oscillation occurs when the contributions to the Zeeman
energy from these two fields become comparable, and the
characteristic length scale of the oscillation is inversely
proportional to the energy gap. Since the gap along $a''$-axis is
much smaller than other directions, the oscillation looks stronger
for $H=(7,0,0)T$ than for other cases.

\begin{figure}[h]
\includegraphics[width=0.33\textwidth,angle=0]{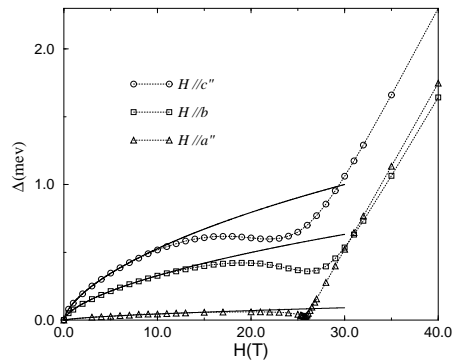}
\vspace{-0.2cm} \caption{Field dependence of the spin excitation
gap. ${\bf D}=(0.13,0.0,0.02)J$. The solid lines are the fitting
curves of the numerical results with Eq. (\ref{delta}) the low
fields. } \label{fig2}
\end{figure}

Figure \ref{fig2} shows the field dependence of the gap
with ${\bf D}=(0.13,0.0,0.02)J$. In low fields, we find that the
gap is well described by the following equation
\begin{equation}
\Delta_\alpha=C^\alpha H^{2/3}\ln^{1/6}(J/\mu_BH_s), \label{delta}
\end{equation}
as predicted by Oshikawa and Affleck. By fitting the numerical
data with the above equation, we find that the coefficient
$C^\alpha $ is given by $C=(0.0072, 0.059, 0.097)$ for ${\bf
D}=(0.13,0.0,0.02)J$. In this system, the gap is finite even in
very low field. This is completely different from the pure
Heisenberg model whose excitations remain gapless until the spin
polarization induced by the applied field becomes saturated. This
difference results from the effective staggered magnetic field
induced by the alternating g-tensor and the DM interaction. Since
this staggered field couples directly to the strongest spin
fluctuation mode at $q=\pi$, it gives rise to the instability of
the ground state and opens a gap in the excitation spectrum. In
high fields, the gap varies almost linearly with $H$ and its slope
is proportional to the corresponding value $g_{\alpha \alpha }^u$.
In the intermediate field regime, the gap varies non-monotonically
with $H$ and shows a local minimum around $H \sim 25T$ in all
three directions. This non-monotonic behavior of the gap is also
due to the competition between the applied field and the induced
staggered field. However, we find that this competition does not
lead to a phase transition in the ground state in this crossover
regime, since the ground state energy drops continuously and
smoothly with $H$.
\begin{figure}[h]
\begin{center}
\includegraphics[width=0.33\textwidth,angle=0]{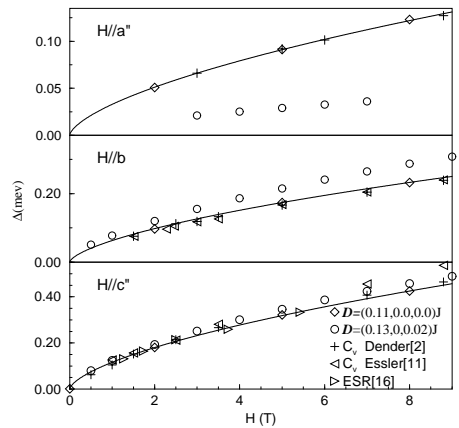}
\end{center}
\vspace{-0.3cm} \caption{Comparison between numerical results
($\circ$ and $\diamond$) and the specific heat $C_v$ ($+$,
$\triangleleft$) and ESR ($\triangleright$) data for the field
dependence of energy gaps. Solid lines are the fitting curves of
the DMRG results to $\Delta_\alpha = C_{\rm DMRG}^\alpha
H^{2/3}\ln^{1/6} (J/\mu_BH_s)$ for ${\bf D}=(0.11, 0.0, 0.0)J$
with $C_{\rm DMRG}=(0.025,0.048,0.089)$. } \vspace{-0.1cm}
\label{figfit}
\end{figure}

Now let us turn to the issue whether the model (\ref{HSDMZ}) is
sufficient to describe the low-energy magnetic properties of
copper benzoate. In the effective field theory, the critical
exponents of the gaps are determined by the relevant and marginal
operators in the low field limit. However, as this theory contains
only the contribution of the dominant effective staggered field
and ignores all other anisotropic terms after eliminating the DM
term, it cannot explain the field-dependence of the gaps
along all three principal axes. Moreover, since the energy gap is
very sensitive to the DM interaction, the value of ${\bf
D}$-vector determined from this theory, ${\bf D} = (0.13, 0.0,
0.02)J$\cite{Osh1}, would also be inaccurate.

To determine accurately the value of ${\bf D}$, we have evaluated
and analyzed the field dependence of the gaps around ${\bf D} =
(0.13, 0.0, 0.02)J$. By comparison with experiments\cite{Den2}, we
find that the gaps for ${\bf D}=(0.11,0.0,0.0)J$ give the best fit
to the experimental data in all three directions (Figure 3). It
also agrees with the recent ESR data for $H//c''$\cite{Nojiri} as
well as the results of Essler for $H//b$ and in low fields for
$H//c''$\cite{Ess}. For this ${\bf D}$ value, our numerical results for the
gap ratios $\Delta_{a''}:\Delta_b:\Delta_{c''}$ in low fields are
$1:1.92:3.56$, in agreement with experiments. However, for ${\bf
D}=(0.13,0.0,0.02)J$, the low field gap ratios are $1:8.2:13.5$
and the theoretical results agree with the experimental data only
along the $c''$-axis (Figure \ref{figfit}). Therefore the
Hamiltonian (\ref{HSDMZ}) is indeed an appropriate model for
copper benzoate and the discrepancy between the effective field
theory and experiments is due to the neglect of the anisotropic
terms rather than due to the interchain coupling as previously
suggested\cite{Osh1}.

For the S=1/2 Heisenberg model, the momentum of the
antiferromagnetic soft mode is shifted from $\pi $ to $\pi \pm
\delta q$ in an applied field. This leads to an incommensurate
peak in the longitudinal structure factor. The momentum shift is
predicted to be proportional to the magnetization:
\begin{equation}
\delta q=2\pi M(H). \label{shift}
\end{equation}
This linear relationship between the momentum shift and the
magnetization was examined by Dender {\it at al} for copper
benzoate\cite{Den2}. They found that the momentum shifts at
$H=3.5, 5, 7T$ for ${\bf H}//b$ were consistent with the
theoretical results for the pure Heisenberg model \cite{Muller}
with $g=2.059$ and $J=1.57$meV. However, as the excitation
spectrum of copper benzoate is fully gaped even in an arbitrarily
small field, unlike in the pure Heisenberg model, this issue needs
also to be reexamined. To do this, we have calculated the spin-spin
correlation functions
\begin{eqnarray*}
S_q^{\alpha \alpha }=\frac1N \sum_{ij}\left\langle \left(
S_i^\alpha -\left\langle S_i^\alpha \right\rangle \right) \left(
S_j^\alpha -\left\langle S_j^\alpha \right\rangle \right)
\right\rangle e^{iq\left( R_i-R_j\right) }.
\end{eqnarray*}
The momentum shift $\delta q$ can be determined from the peak
position of $S_q^{\alpha \alpha }$. From our calculations, we find
that $S_q^{\alpha \alpha }$ behaves differently along the three
principal axes. $S_{q}^{bb}$ shows a weak incommensurate peak.
However the peaks of $S_{q}^{a''a''}$ and $S_q^{c''c''}$ are
pinned at $q=\pi$. With $H$ increasing, the heights of these
peaks are suppressed in all three directions.

\begin{figure}[h]
\includegraphics[width=0.33\textwidth,angle=0]{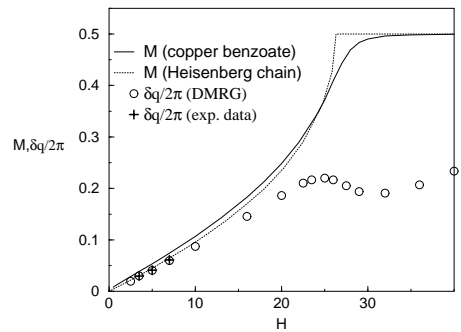}
\caption{Magnetization $M$ versus the applied field $H$
 with ${\bf D}=(0.11,0,0)J$ (dashed line) and the
pure Heisenberg model (solid line) for $H//b$.
The corresponding shift $\delta q/2\pi$ ($\circ$) is
shown and also compared with the experimental
data ($+$) \cite{Den2}.} \label{figmagu}
\end{figure}

Figure (\ref{figmagu}) shows the field dependence of the
magnetization $M(H)$ and the momentum shift $\delta q$ for ${\bf H}//b$.
For comparison, the magnetization curve for the pure Heisenberg
model and the experimental data for $\delta q$ are also shown in
the figure. As seen from the figure, our numerical values of
$\delta q$ agree quantitatively with the experimental data
\cite{Den2}. However, we find that Eq. (\ref{shift}) is not valid
for copper benzoate. In particular, in contrast to $M(H)$, $\delta q$
 varies non-monotonically with $H$.  We believe this
non-monotonic variation of $\delta q$ with $H$ can be detected by
neutron scattering measurements. In addition, the magnetization is
gradually saturated for large fields as a direct consequence of
the DM interaction. This is different from in the pure Heisenberg
model where the spins are fully polarized beyond a critical field
$\mu_B H_c=J$, at which the two magnetization curves differ
mostly.

\begin{figure}[h]
\includegraphics[width=0.33\textwidth,angle=0]{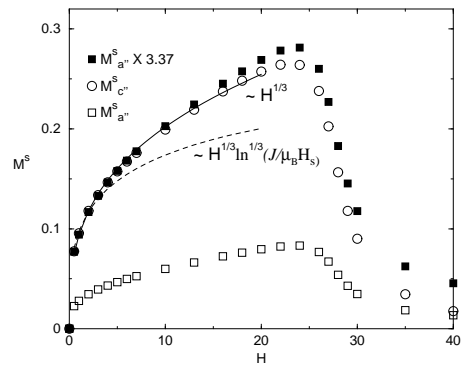}
\caption{The staggered magnetization as a function of $H$ for
${\bf H}//b$ and ${\bf D}=(0.11,0,0)J$} \label{figM_s}
\end{figure}

When ${\bf H}//b$, spins are polarized ferromagnetically along the
$b-$axis and antiferromagnetically along the other two directions.
Figure (\ref{figM_s}) shows the staggered magnetization as a
function of $H$ for ${\bf D}=(0.11,0,0)J$. At low fields,
both $M^s_{a''}$ and $M^s_{c''}$ behave as $H^{1/3}$. This field
dependence of the staggered magnetization above $2T$ deviates from
the $H^{1/3}\ln^{1/3} (J/\mu_BH_s)$ behavior predicted by the
effective field theory \cite{Osh1} and is not sensitive to the
{\bf D}-vector. It may well be that the range of the validity of
the leading approximation in the effective field theory is
narrower for the staggered magnetization than that for the gap. We
should also mention that the numerical accuracy is lower when $H$
is below 0.5T. In intermediate fields, both $M^s_{a''}$ and
$M^s_{c''}$ vary non-monotonically with $H$, in analogy with the
field dependence of energy gaps. In the large $H$ limit, $M_s$
tends to approach zero. This means that the staggered field $H_s$
induced by the external field is overwhelmed by the uniform field
$H$ in the high field limit, although $H_s$ is approximately
proportional to $H$.

\begin{figure}[h]
\includegraphics[width=0.34\textwidth,angle=0]{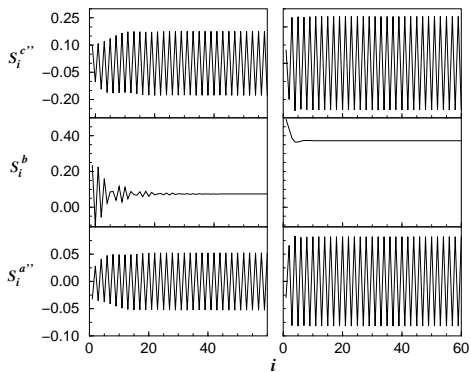}
\vspace{-0.2cm}
\caption{Local spin polarization for ${\bf
H}=(0,7,0)T$ (left) and ${\bf H}=(0,25,0)T$ (right) for $L=120$
and ${\bf D}=(0.11,0,0)J$. $S_i^b$ is symmetric with respect to
the middle point of chain, while the others are antisymmetric.
Only is half of the chain shown.} \label{figcomp}
\end{figure}

The above results show that the DM interaction affects strongly
the properties of low-lying excitations although this term is much
smaller than the Heisenberg exchange interaction. In particular,
the staggered magnetic field induced by this term opens a gap in
the spin excitations and leads to an antiferromagnetic long range
order to accompany the ferromagnetic long range order induced by
the applied field, shown in Fig.\ref{figcomp}. The coexistence of
orthogonal ferromagnetic and antiferromagnetic long range orders
in an external field is a characteristic feature of quantum spin
systems with DM interactions.

In conclusion, we have shown that the microscopic model (1)
including two inequivalent lattice sites and the DM interaction is
a very good description of available experimental data. It can
reproduce correctly the gap values in all three  directions as
well as the shift of the incommensurate peaks, without resorting
to any additional interaction like interchain coupling. The
crossover to strong field behavior and the breakdown of the
relation between the momentum shift and the magnetization call for
further experimental studies. Finally, we would point out that
although the DM interaction is generally small, its effects could
be very important for real materials. In particular, for metallic
materials, e.g. Yb$_4$As$_3$\cite{Kohgi}, one could use external
fields to modulate the magnetic transport properties which is one
of the focal points for the spintronics.

We wish to thank D.C. Dender and C. Broholm for correspondence and
I. Affleck, C. F. Chen, F. Essler, H. Nojiri, M. Oshikawa and K. Ueda for
fruitful discussions. This work is supported in part by the
Special Funds for Major State Basic Research Projects of China and
by the National Natural Science Foundation of China.


\vfill
\end{document}